# Phase separated behavior in Yttrium doped $CaMnO_3$

Neetika Sharma[1], A. Das[1,a], C.L. Prajapat[2], Amit Kumar[1], M.R. Singh[2]

[1]Solid State Physics Division, Bhabha Atomic Research Centre,
Trombay, Mumbai 400085, India

[2]Technical Physics Division, Bhabha Atomic Research Centre,
Trombay, Mumbai 400085, India

**Abstract**

The effect of electron doping on the structural, transport, and magnetic properties of Mn (IV) - rich $Ca_{1-x}Y_xMnO_3$ ($x \leq 0.2$) samples have been investigated using neutron diffraction, neutron depolarization, magnetization and resistivity techniques. The temperature dependence of resistivity follows the small polaron model and the activation energy exhibits a minimum for x=0.1 sample. A phase separated magnetic ground state consisting of ferromagnetic domains (~7μm) embedded in G-type antiferromagnetic matrix is observed in the sample, x = 0.1. The transition to the long-range magnetically ordered state in this sample is preceded by a Griffith's phase. On lowering temperature below 300K a structural transition from orthorhombic structure (*Pnma*) to a monoclinic structure (*$P2_1/m$*) is observed in the case of x=0.2 sample. The ferromagnetic behavior in this case is suppressed and the antiferromagnetic ordering is described by coexisting C-type and G-type magnetic structures corresponding to the monoclinic and orthorhombic phases, respectively.

a) Corresponding author

   Amitabh Das
   Solid State Physics Division
   Bhabha Atomic Research Centre
   Mumbai 400085, India
   Email address   adas@barc.gov.in

**Introduction**

The hole doped manganites $A_xA'_{1-x}MnO_3$ exhibit several interesting behavior *viz*., colossal magnetoresistance (CMR), charge, spin and orbital coupling, and mesoscopic phase separation. A strong correlation between the charge, spin, and orbital degrees of freedom in these compounds makes them sensitive to external perturbations, such as temperature, magnetic field, external pressure, and average A-site ionic radii [1, 2]. The physical properties of these compounds are mostly governed by Zener double-exchange mechanism of electron hopping, superexchange interactions, and Jahn-Teller type electron-phonon interactions [3]. The effect of hole doping in the magnanites has been studied in great detail but only a few studies are available on the effect of electron doping on their magnetic properties.

The varied magnetic structures observed in the series $La_{1-x}Ca_xMnO_3$ system have been described by Wollan and Koehler [4]. $CaMnO_3$ crystallizes in the perovskite related $GdFeO_3$-type orthorhombic structure and exhibits a G-type antiferromagnetic (AFM) structure ($T_N$ ~125 K) with a weak ferromagnetic component in its ground state [5]. In this structure each Mn moment is coupled antiferromagnetically with its nearest Mn neighbors. However, doping with trivalent ions at $Ca^{2+}$ site is found to give rise to ferromagnetic (FM) behavior in these compounds. The role of spin canting and/or phase separation leading to FM behavior in these compounds has been a subject of debate in the literature. The ferromagnetic behavior in the series of compounds $Ca_{1-x}La_xMnO_3$ has been shown to be intrinsic in origin and is found to couple strongly with the lattice leading to a complex structural and magnetic phase diagram [6, 7]. The transport and magnetic properties of some of the electron doped manganites $Ca_{1-x}A_xMnO_3$ (A= Pr, Nd, Eu, Gd, Ho, Sm, Ce , Th) are found to be governed by the electron concentration [8, 9] as against the dominating influence of average A-site ionic radii observed in the case of hole doped manganites. Phase separation behavior has been reported for Sm [9] and Pr [10] doped $CaMnO_3$ whereas the magnetic ground for the Ho [11] doped compound is identified as a spin canted antiferromagnet. Similar studies on Ru doped $CaMnO_3$ show phase separated FM + AFM ground state [12]. However, the antiferromagnetic structure of Mo doped $CaMnO_3$ was found to be $A_XF_yG_Z$ type and a clear distinction between phase separation and spin canting behavior in this compound could not be established [13].

These studies therefore, show that the nature of magnetic ordering in electron doped manganites is varied and inconclusive. Doping $CaMnO_3$ with $Y^{3+}$ ion which is non magnetic and similar to $La^{3+}$, albeit with lower ionic radii, it was expected, that an equally complex magnetic phase diagram would emerge [14,15]. However, a previous study on Y doped $CaMnO_3$ could not establish the presence of either phase-separation or homogeneous canted AFM magnetic structure [14]. We have investigated the isostructural $Ca_{1-x}Y_xMnO_3$ ($0 \leq x \leq 0.2$) compounds and show that a phase separated behavior, with coexisting short range ferro- and long range antiferromagnetic ordering describes the magnetic state of $x = 0.1$ compound. At higher doping ($x = 0.2$), the orthorhombic phase partially transforms to a monoclinic phase. Antiferromagnetic ordering of type $G_z$ for the orthorhombic structure and C-type ordering for the monoclinic structure is observed in this case. Our experimental results are not in agreement with the recently concluded spin canting behavior observed theoretically [16] and experimentally in Ce-doped $CaMnO_3$ [17].

**Experimental Details**

Polycrystalline samples of $Ca_{1-x}Y_xMnO_3$ (x = 0.1, 0.2) were synthesized using conventional solid-state reaction methods. The starting material $CaCO_3$, $MnO_2$ and $Y_2O_3$ were mixed in stoichiometric ratios and heated in air at 1100°C for 30h, 1150°C for 20h, and 1300°C for 30h, successively with intermediate grindings. X-ray powder diffraction patterns were recorded using Cu Kα radiation in the angular range $10° \leq 2\theta \leq 70°$ on a Rigaku make diffractometer. The dc resistivity measurements were carried out using standard four probe technique. The magnetization measurements were recorded on a SQUID magnetometer (Quantum Design). Neutron depolarization measurements (λ = 1.205Å) were carried out on the polarized neutron spectrometer at Dhruva reactor, Bhabha Atomic Research Centre, Mumbai, India, with $Cu_2MnAl$ (1 1 1) as polarizer and $Co_{0.92}Fe_{0.08}$ (2 0 0) as analyzer. Neutron diffraction patterns were recorded on the PD2 powder diffractometer (λ = 1.2443Å) at the Dhruva reactor, Bhabha Atomic Research Centre, Mumbai in the angular range $5° \leq 2\theta \leq 140°$. The Rietveld refinement of the neutron diffraction patterns were carried out using FULLPROF program [18].

**Results and Discussion**

The Rietveld refinement of the room temperature x-ray diffraction pattern confirms the single – phase nature of the studied samples $Ca_{1-x}Y_xMnO_3$ ($0 \leq x \leq 0.2$) and is shown in figure 1. All the samples crystallize in the orthorhombic phase (space group *Pnma*) at 300K. The cell parameters of $Ca_{0.9}Y_{0.1}MnO_3$ sample follow the relation $a > b/\sqrt{2} > c$ in the temperature range $6K \leq T \leq 300K$, which corresponds to an O-type orthorhombic structure. This structure results from a cooperative buckling of the corner shared octahedra [19]. No evidence of monoclinic phase is found in this compound on lowering of temperature. Therefore, we have analyzed the neutron diffraction data of x = 0.1 sample in the *Pnma* space group alone. However, analysis of the x-ray and neutron diffraction data at 12K for $Ca_{0.8}Y_{0.2}MnO_3$ sample shows that this sample exhibits a monoclinic phase (*P2₁/m* space group) in addition to the orthorhombic phase. Clear distinct reflections arising from the monoclinic phase are observed in the x-ray diffraction pattern recorded at 12K for x = 0.2 (figure 1). Therefore, a two phase refinement, including both orthorhombic and monoclinic phases has been carried out for this sample. At 6K the monoclinic phase is found to be dominant with 82% volume fraction. The fraction of monoclinic phase gradually decreases with increase in temperature and at 300K the fraction of monoclinic phase is very small (~ 6%), as shown in the inset of figure 1. In this sample, the orthorhombic cell parameters exhibits a > b/√2 > c at 300K. The unit cell volume, increases with Y substitution despite the lower ionic radius of $Y^{3+}$ (1.07 Å) as compared to $Ca^{2+}$ (1.18 Å). The observed increase in volume, therefore, results from the larger ionic radius of $Mn^{3+}$ (0.645 Å) as compared to $Mn^{4+}$ (0.530 Å) in six coordinated state which compensates for the difference of ionic radii between $Ca^{2+}$ and $Y^{3+}$ ions [20]. The difference in ionic radii of $Ca^{2+}$ and $Y^{3+}$ leads to ionic-size disorder which is quantified by the A-cation radius distribution expressed as $\sigma^2 = \sum x_i r_i^2 - \langle r \rangle^2$, where $x_i$ is the fractional occupancy of the A-site ion, $r_i$ is the corresponding ionic radius and $\langle r \rangle$ is the average A-site ionic radius [21, 22]. The $\langle r \rangle$ decreases with increase of the $Y^{3+}$ content because of the lower ionic size of Y ion. This reduction in $\langle r \rangle$ induces a tilt of the $MnO_6$ octahedra which results in the localization and ordering of $Mn^{3+}/Mn^{4+}$ cations. The disorder $\sigma^2$ and lattice distortion parameter (D) values are given in the Table 1. Previously, the structural changes in Y doped $CaMnO_3$ have been reported by Vega et al [15] from analysis of x-ray diffraction patterns at 300K. They show the structure remains O- orthorhombic for x ≤ 0.25, O + O′ orthorhombic in the region 0.25 < x < 0.5, and O′ orthorhombic for 0.5 ≤ x ≤ 0.75. Our results are in partial agreement with the previous studies where we find for x≤0.1 sample the structure remains O-type orthorhombic (*Pnma* space group) in the whole temperature range while a transition from orthorhombic phase to monoclinic phase is observed in x=0.2 sample on lowering of temperature. Similar transition to monoclinic phase has been reported in some of the earlier studies on electron doped manganites [23 - 25]. The presence of the monoclinic phase has been correlated with the nature of magnetic ordering [6, 7, 23].

In our previous study on $CaMnO_3$, we had reported the value of resistivity ($\rho_{300K}$) at 300K for $CaMnO_3$ ~1.7Ω cm [26]. On doping with Y, the value of $\rho_{300K}$ decreases (Table 1). This behavior of resistivity is consistent with earlier studies on effect of Y doping reported by Aliaga et al and Sudheendra et al. [14, 27]. The resistivity values at 300K for these samples are found to decrease with increase in $Y^{3+}$ concentration. However, the doping does not lead to a metal-insulator transition at low temperatures. When $Y^{3+}$ is substituted for $Ca^{2+}$ in $CaMnO_3$, $Mn^{3+}$ ions are introduced in the system as a result of charge compensation. The decrease in the value of resistivity in Y doped samples has been attributed to the presence of $Mn^{3+}$ ions which introduces ferromagnetic double exchange between $Mn^{3+}$ and $Mn^{4+}$ ions. Similar decrease in the resistivity with doping by a trivalent ion at A-site has been previously reported in $Ca_{1-x}A_xMnO_3$ (A= La, Gd, Nd, Tb, Ho) systems [8, 9, 11, 28]. An insulator to metal transition has been observed in $Ho^{3+}$ and $Ce^{4+}$ doped $CaMnO_3$ albeit, in the high temperature region (300-900K) [28, 29]. The temperature dependence of resistivity of $Ca_{1-x}Y_xMnO_3$ samples is shown in figure 2(a). Semiconductor like behavior is observed in both the samples. In the case of x=0.1 sample a departure from semiconducting behavior is observed for T < 125K which coincides with the magnetic ordering temperature of this sample and therefore, this behavior may be correlated with the magnetic nature of the sample. Short range ferromagnetic correlations are observed for x=0.1 sample in the magnetically ordered (T < $T_N$) region as well as in the paramagnetic region (discussed later) which possibly explains the reduction in resistivity, $\rho(0)$. However, the compound continues to remain insulating indicating that the size of the FM clusters in the magnetically ordered state are below the percolation threshold. The temperature dependence of resistivity in the temperature region ($T_N \leq T \leq 300K$) is found to follow small polaron model as shown in figure 2(b). The temperature variation of resistivity due to small polarons is expressed as $\rho(T) = AT \exp(W/k_B T)$, where W represents the activation energy. The variation of W and resistivity at 50K ($\rho_{50K}$) with x is shown in figure 3. A significant reduction in both the activation energy and $\rho_{50K}$ is observed in the case of x=0.1 sample and is attributed to the presence of ferromagnetic clusters in this sample. With further increase in Y concentration, the value of activation energy increases from 29meV for x=0.1 sample to 61meV for x=0.2 sample as shown in figure 3. The increase in the value of activation energy for x > 0.1 could be related to the increase in tendency to form small polarons when the concentration of $Y^{3+}$ increases (concentration of $Mn^{3+}$ ions increases) [30]. Similar transport behavior has been previously reported in Sb doped $CaMnO_3$ where the value of activation energy is found to increase gradually with increase in Sb concentration in spite of a monotonous decrease in the resistivity in the high temperature region [31].

Figure 4 shows the temperature dependence of magnetic susceptibility for $Ca_{1-x}Y_xMnO_3$ (x = 0, 0.1, 0.2) samples. For temperature below 110K, a significant enhancement in the magnetic susceptibility is observed for $Ca_{0.9}Y_{0.1}MnO_3$, indicating the presence of a magnetic transition with FM component. A strong ferromagnetic moment of ~1.2$\mu_B$ is observed for the x=0.1 sample as shown in the M(H) curve in the inset of figure 4. The inset (b) of figure 4 shows the inverse of susceptibility ($\chi^{-1}$) as a function of temperature for the samples (x = 0.1, 0.2). In the paramagnetic region, the inverse susceptibility follows the Curie-Weiss law, given by $\chi = C/(T-\theta)$, where $\chi$, C and $\theta$ are the magnetic susceptibility, Curie constant and Curie-Weiss temperature, respectively. However, for the x=0.1 sample, $\chi^{-1}$ (T) shows a deviation from Curie-Weiss law in the paramagnetic region. The deviation of $\chi^{-1}$ from a linear dependence with temperature from well above transition temperature is a characteristic feature of Griffiths phase [32-34]. The temperature dependence of susceptibility in Griffiths model is given by $\chi^{-1} = (T - T_C)^{1-\lambda}$, where λ is the susceptibility exponent and varies between 0 and 1. A fit to Griffiths model is shown in the inset of figure 4 with λ ~ 0.12. The presence of Griffiths like phase in x=0.1 sample, suggests the presence of short range ferromagnetic (FM) clusters in the paramagnetic region. The Griffiths like behavior has been observed in various systems like intermetallics, hole doped manganites, layered manganites and in few electron doped manganites [35-38]. The emergence of Griffiths phase is indicative of the presence of competing ferromagnetic double exchange and antiferromagnetic superexchange interactions in the system. A fit of the inverse susceptitbility data, well above the respective transition temperatures (150K < T 300K), to the Curie – Weiss behavior yields a negative Curie temperature of -69K and -96K for samples x =0.1 and 0.2, respectively (Table 1). The negative value of the Curie temperature indicates antiferromagnetic interactions in the system. This is in agreement with the neutron diffraction studies where we find antiferromagnetic ordering in both these compositions. In x=0.1, the magnetic structure is of $G_Z$ type (100K < $T_N$ < 125K) while in the case of x=0.2 a mixture of C-type (125K < $T_N$(C) < 150K) and G-type ($T_N$~100K) magnetic ordering corresponding to the two crystallographic phases is observed (discussed later). In a previous study of $Ca_{1-x}Y_xMnO_3$ compounds, positive value for Curie temperature was reported for x=0.1, 0.2, indicating FM interactions in the system [14]. From the values of Curie constant, we have calculated the effective paramagnetic moment ($\mu_{eff}$). As a result of trivalent doping, the Mn site is occupied by a mixture of $Mn^{4+}$ and $Mn^{3+}$ ions. Therefore, $\mu_{eff}$ is calculated as $\mu_{eff}^{cal} = \sqrt{x\mu_{eff}^2(Mn^{3+}) + (1-x)\mu_{eff}^2(Mn^{4+})}$, where, $\mu_{eff}$ for $Mn^{4+}$ (S = 3/2) and for $Mn^{3+}$ ( S = 2 ) is 3.87 $\mu_B$ and 4.89 $\mu_B$, respectively. The expected values of $\mu_{eff}$ are 3.99 $\mu_B$ and 4.1$\mu_B$, which is close to the experimentally obtained values of 4.13 $\mu_B$ and 4.65 $\mu_B$ for samples x = 0.1 and 0.2, respectively.

**Magnetic structure**

Neutron diffraction patterns for both the samples have been recorded at various temperatures between 6K and 300K. In the case of x=0.1 sample, superlattice reflections at low angles are observed on lowering of temperature below 125K. These reflections could be indexed with $\vec{k}$=0 propagation vector in $P\bar{1}$ space group. The strong enhancement in the intensity of (011) reflection indicate that the intensity is of magnetic in origin as the nuclear contribution is absent in this reflection. Additionally, we do not observe any enhancement in the intensity of the fundamental Bragg reflections. This indicates the absence of long range ferromagnetic component in this sample. The basis vectors have been determined by using BASIREPS program with propagation vector $\vec{k}$ =0. To represent the magnetic structure of these compounds Bertaut's notation in the *Pnma* setting has been adopted. For x = 0.1 sample, at 6K the magnetic structure is found to be $G_Z$ ($\Gamma_4$) type with the spins coupled antiferromagnetically along the z-axis (crystallographic c-axis). The moment on Mn is found to be 2.1$\mu_B$ at 6K, which is lower than the expected value of 3.1$\mu_B$ for this compound. The thermal variation of the magnetic moment exhibits a Brillioun-type temperature dependence and is shown in the inset of figure 5. However, the temperature dependence of susceptibility (figure 4), shows a rapid increase below 110K, indicating the presence of ferromagnetic correlations in x=0.1 sample. The presence of ferromagnetic behavior in magnetization studies of $Ca_{0.9}Y_{0.1}MnO_3$ raises an important question whether there is a canting of the G-type AFM structure, as originally proposed by deGennes [39], or a coexistence of magnetic phase separation to form domains of AFM and FM ordering. The absence of any magnetic contribution in the form of enhancement in the intensity of the low angle fundamental reflections in our neutron diffraction results rules out the presence of canted AFM. The magnetic structure of this compound is similar to the $G_Z$-type magnetic ordering we had reported earlier in the parent compound, $CaMnO_3$ [26]. However, the site moment $2.1\mu_B$ for x=0.1 is lower than $2.84 \mu_B$ for x=0 sample (expected value of $3\mu_B$ for $Mn^{4+}$). Similar low value of moments has been observed in other doped samples [6, 7, 17] and attributed to the Mn (3d) - O(2p) hybridization [9]. It could also be an indication of incomplete ordering of the Mn ions, below the $T_N$, in this sample.

Figure 6 shows a section of the neutron diffraction data at 300K and 6K for $Ca_{0.8}Y_{0.2}MnO_3$. As shown in the inset of figure 1, this sample exhibits a structural transition to a monoclinic phase on cooling below 300K and both these phases coexist over a large temperature region. For T<125K, superlattice reflections in addition to those corresponding to G-type magnetic ordering are observed. This strong reflection could be indexed as (100) in the $P2_1/m$ space group. The neutron diffraction pattern has been analyzed taking into account both the monoclinic (*$P2_1/m$* space group) and orthorhombic (*Pnma* space group) phases. The basis vectors have been determined by using BASIREPS program with propagation vector $\vec{k} = (\frac{1}{2}\ 0\ \frac{1}{2})$ for *$P2_1/m$* space group and $\vec{k}= (0\ 0\ 0)$ for *Pnma* space group. The magnetic structure is found to be C-type AFM corresponding to monoclinic phase and G-type AFM for orthorhombic phase. The value of the moment on Mn at 6K is found to be $2\mu_B$ for both the phases, which is lower than the expected value of $3.2\mu_B$. The lower value of moments in this sample too indicates that part of the Mn moments may not be completely ordered below $T_N$. The C-AFM structure is characterized by a ferromagnetic ordering of the magnetic moments of Mn ions in chains and by an antiferromagnetic coupling between neighboring chains suggesting double-exchange interactions along the chains and superexchange interactions (SE) between neighboring chains. Therefore, in the presence of C-type AFM, 1D FM correlations leads to the $d_{3z^2-r^2}$ orbital polarization and hence it further leads to lowering of the symmetry from orthorhombic (*Pnma*) to monoclinic (*P2$_1$/m*) [6]. The $T_N$ (C-type) is found to be higher (125K <$T_N$(C) <150K) than $T_N$ (G-type) (~100K). Similar behavior has been found in the case of $Ca_{1-x}La_xMnO_3$ (0 ≤ x ≤ 0.2) [6]. However, there is no evidence of ferromagnetic behavior in this sample. In earlier reported studies on single electron doped $Ca_{1-x}A_xMnO_3$ systems, a canted G-type structure was observed only up to x~0.1, which then changed to C-type structure [6, 24, 40-43] for higher values of x, which is similar to our results on Y doped $CaMnO_3$. Similar, crystallographic and magnetic phase diagram has been observed in case of two electron doped $Ca_{1-x}Ce_xMnO_3$ (x≥0.075) system [17]. Eventually, the studied system tends to stabilize in the C-type AFM structure which comprises of FM chains. The absence of ferromagnetic behavior has also been observed in the case of $Ca_{0.85}Sm_{0.15}MnO_3$ [44]. The low temperature crystallographic and magnetic phase, in this compound is described by C-P2$_1$/m and G-Pnma. However, application of magnetic field is found to result in F-Pnma phase together with G-Pnma. This result therefore, bridges our observations between 10% and 20% doping levels. The evolution of the long range magnetic ordering in electron doped systems is supported by previously published theoretical work [1, 45, 46]. Theoretically it has been observed that the systematics of the phase diagram in manganites changes considerably as a function of $J_H$ (Hund's coupling), $J_{AF}$ (superexchange coupling) and (W) band width. Pai et al. has summarized the magnetic phase diagram for electron doped manganites for different electron concentration as a function of $J_H$ and $J_{AF}$ [47]. For low electron concentration, the SE (superexchange) interaction dominates over Hund's coupling and leads to G-type structure. With further electron doping, the kinetic energy starts dominating over the SE coupling. The competition between effective kinetic energy (determined by $J_H$) and SE leads to transition from G-type to C-type. The electron doping concentration and competition between $J_H$ and $J_{AF}$ leads to different magnetic structures (G-C-A-F).

**Neutron depolarization studies**

Unlike, in the case of x = 0.2 compound where no evidence of ferromagnetic correlations are observed, in x = 0.1 we had observed that the magnetic susceptibility increases rapidly below 110K, indicating the presence of ferromagnetic correlations. However, it was inconclusive from neutron diffraction experiments. Therefore, this behavior has been further studied using neutron depolarization measurements. In this experiment, we measured the flipping ratio R (ratio of transmitted intensities for two spin states of the incident neutron spin) which is a measure of the transmitted beam polarization. The flipping ratio is expressed in the form $R = \frac{1 - P_i D P_A}{1 + (2f - 1) P_i D P_A}$ [48] where, $P_i$ is the incident beam polarization, $P_A$ is the efficiency of the analyzer crystal, f is the efficiency of the DC flipper and D is the depolarization coefficient (due to the sample under investigation). Figure 7 shows the temperature dependence of flipping ratio for x = 0.1 and 0.2 samples. For x=0.1 sample, flipping ratio remain constant up to ~ 110K, below which, the flipping ratio decreases rapidly, reaching a minimum at ~ 90K and after that remains constant down to 2K. This decrease in flipping ratio below 110K indicates the presence of ferromagnetic correlations which correlates well with our magnetization study on this compound. An estimate of the domain size in the ferromagnetic region is obtained using the expression $P_f = P_i \exp[-\alpha(d/\delta)\langle\phi_\delta\rangle^2]$, where $P_f$ and $P_i$ are the transmitted beam and incident beam polarization, respectively, α is a dimensionless parameter (=1/3), d is the sample thickness, δ is a typical domain length and the precession angle $\phi_\delta = (4.63 \times 10^{-10} Oe^{-1} A^{-2}) \lambda \delta B$. The domain magnetization, B is obtained from the bulk magnetization. This expression is valid in the limit where domains are randomly oriented and the Larmor precession of the neutron spin due to the internal magnetic field of sample is small fraction of 2π, over the typical domain length scale [49, 50]. The estimated domain size for x=0.1 sample is ~ 7 µm. This measurement therefore, gives a clear evidence of the existence of the ferromagnetic domains. Similar ferromagnetic clusters in antiferromagnetic phase have been observed in LPCMO system by Uehara et al [51] and more recently observed in intermediate compositions in doped $CaMn_{1-x}W_xO_3$ (0≤x≤0.1) [52], $CaMn_{1-x}Mo_xO_3$ [53] $Ca_{0.85}Pr_{0.15}MnO_3$ [54]. In the case of x = 0.2, the flipping ratio remain constant down to the lowest temperature indicating the absence of ferromagnetic domains in this sample. In an antiferromagnet, since there is no net magnetization, depolarization is not expected. This is in agreement with the magnetization and neutron diffraction studies on this sample, where no evidence of enhancement in the intensity of fundamental Bragg reflections (101) (020) is observed. Since neutron depolarization measurements provide information on the magnetic inhomogeneity on a length scale > 1µm we rule out the presence of ferromagnetic correlation at this length scale in x= 0.2 sample.

Monte Carlo simulations on the manganites have shown the existence of large FM clusters in the antiferromagnetic phase, when the densities of ferromagnetic and antiferromagnetic phases are equal [55, 1]. The variation in the size of clusters has been explained by Moreo et al. [55] on the basis of disorder in the system. They find that introducing disorder into AFM matrix leads to growth of FM clusters. The size of the clusters eventually shrinks on increasing the disorder. We attribute the absence of ferromagnetic clusters (of micrometer length scale) in x=0.2 to increase in disorder. For x=0.1 sample, the magnetic ground state is explained by considering phase separated state, which consists of short range FM clusters embedded in the G-type AFM matrix. Similar behavior has been observed before in the case of half doped manganites [56]. Its presence at different length scales in doped manganites has been discussed by Shenoy et al. [57]. Phase separation (coexistence of FM and AFM) behavior has been observed experimentally in single electron doped systems such as $Ca_{1-x}La_xMnO_3$ for $0<x\leq0.09$, $Ca_{1-x}Pr_xMnO_3$ for $x\leq0.1$, and $Ca_{0.9}Tb_{0.1}MnO_3$ [58]. However, in the two electron doped $Ca_{1-x}Ce_xMnO_3$ for $0<x\leq0.075$ system, the magnetic ground state is better described by canted AFM model. The origin of weak ferromagnetism in doped manganites, in terms of spin canted state was originally proposed by deGennes [39]. However, later theoretical studies have shown that the phase separated state in doped manganites is the more stable phase [44, 59-62]. We attribute the phase separated behavior with varying coexisting magnetic structures in the studied $Ca_{1-x}Y_xMnO_3$ to the disorder in the system.

**Conclusion**

We have investigated the magnetic, transport and structural properties of polycrystalline $Ca_{1-x}Y_xMnO_3$ ($x\leq0.2$). The compounds crystallize in orthorhombic structure for $x\leq0.1$ while a coexistence of monoclinic and orthorhombic phase is observed at low temperatures in the case of $x=0.2$. The temperature dependence of the resistivity exhibits a semiconducting behavior and is described by small polaron model. The activation energy exhibits a minimum at $x=0.1$ and is ascribed to presence of ferromagnetic clusters in this sample. The magnetic structure for $x=0.1$ sample, at 6K is found to be $G_Z$ type similar to the parent compound. The rapid increase in magnetization and decrease in flipping ratio below 110K, indicates the presence of ferromagnetic correlations (~7μm). These studies together indicate a phase separated state with FM clusters embedded in the AFM matrix in $x=0.1$ sample. Neutron diffraction and neutron depolarization studies rule out the presence of ferromagnetic correlations in x=0.2 sample. For x=0.2, the orthorhombic phase partially transforms to a monoclinic phase. An antiferromagnetic ordering of the type $G_z$ for the orthorhombic phase and C-type ordering for the monoclinic phase is observed in x=0.2 sample.

**Figure captions:**

**Figure1:** X- ray diffraction patterns of $Ca_{1-x}Y_xMnO_3$ ($0 \leq x \leq 0.2$) at 300K and x=0.2 at 12K. Open circles are observed data points. The solid line represents the Rietveld refinement. The tick marks indicates the position of nuclear Bragg peaks. In the case of x=0.2 sample the upper and lower tick marks indicate the position of reflections in Orthorhombic and monoclinic phase, respectively. The plots for x=0.1 and 0.2 are offset vertically for clarity. Inset shows the temperature variation of the monoclinic and orthorhombic phase fraction for x=0.2. The solid lines are the guide to the eye.

**Figure2**: (a): Electrical resistivity ($\rho$) versus temperature ($T$) for $Ca_{1-x}Y_xMnO_3$ ( x = 0.0, 0.1, 0.2). (b) The variation of ln ($\rho/T$) with inverse of temperature. The solid line through the data is the fit to the small polaron model.

**Figure 3**: The variation of activation energy (W) and resistivity at 50K ($\rho_{50K}$) as a function of composition. The solid line is a guide to the eye.

**Figure4**: The temperature variation of the magnetic susceptibility for $Ca_{1-x}Y_xMnO_3$ ($0 \leq x \leq 0.2$). Inset (a) shows the variation of M with magnetic field at T = 5K. Inset (b) shows the inverse of susceptibility as a function of temperature and fit to Griffiths model (x=0.1) and Curie- Weiss law (x=0.2).

**Figure5**: The observed (symbols) and calculated (line) neutron diffraction pattern for $Ca_{0.9}Y_{0.1}MnO_3$ compound at T = 6 K. Lower solid line is the difference between observed and calculated pattern. The first row of tick marks indicates the position of nuclear Bragg peaks and second row indicate the position of magnetic Bragg peaks. Inset shows the thermal variation of magnetic moments. The obtained magnetic structure is drawn in the inset.

**Figure6**: The observed (symbols) and calculated (line) neutron diffraction pattern for $Ca_{0.8}Y_{0.2}MnO_3$ compound at T = 6K and 300K. Lower solid line is the difference between observed and calculated pattern. The first and second row of tick marks indicates the position of nuclear Bragg peaks for *$P2_1/m$* and *Pnma* space group, respectively. The third and fourth row indicates the position of magnetic Bragg peaks for C-type and G-type AFM, respectively. Inset shows the thermal variation of magnetic moments. The obtained magnetic structures are drawn in the inset.

**Figure7**: The flipping ratio (R) versus temperature for $Ca_{1-x}Y_xMnO_3$ (x=0.1, 0.2)

**Figure 1**

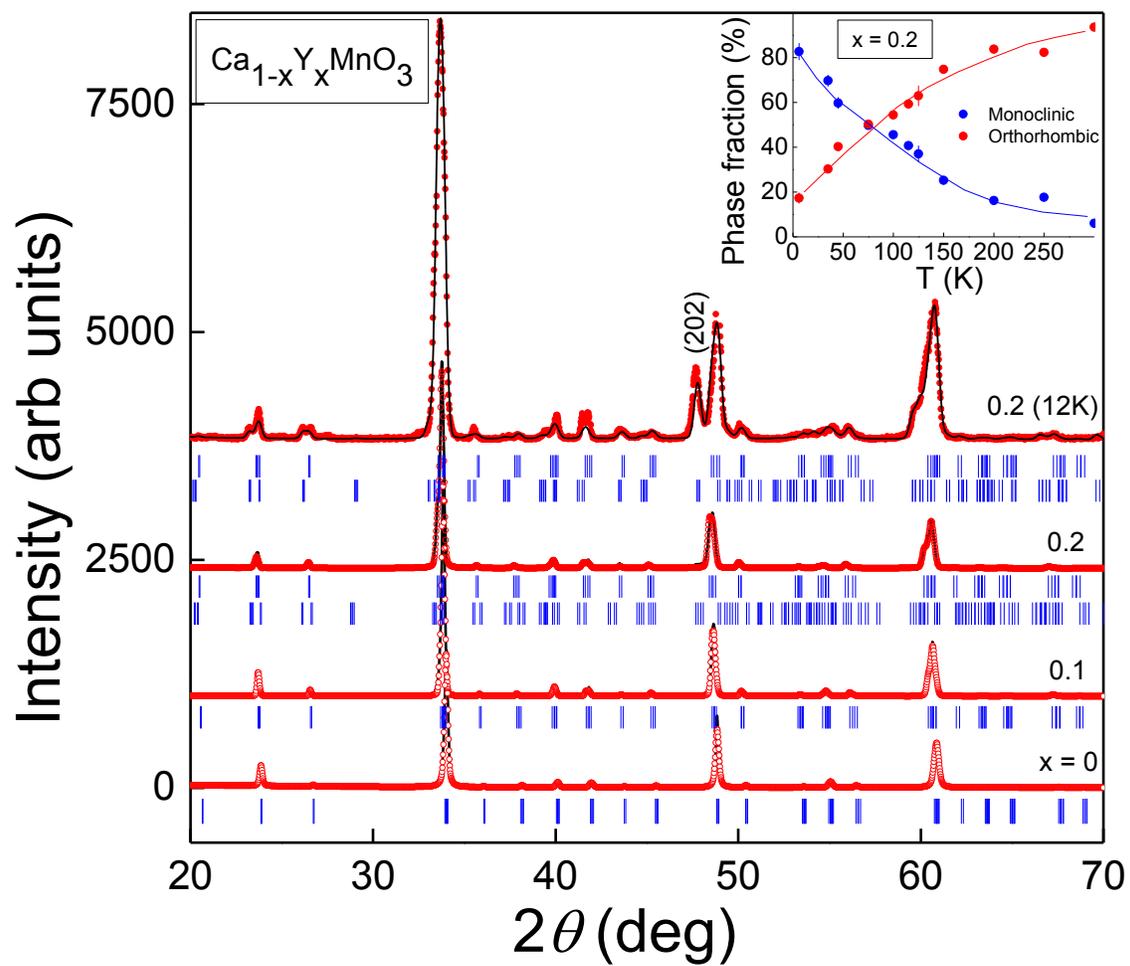

**Figure 2**

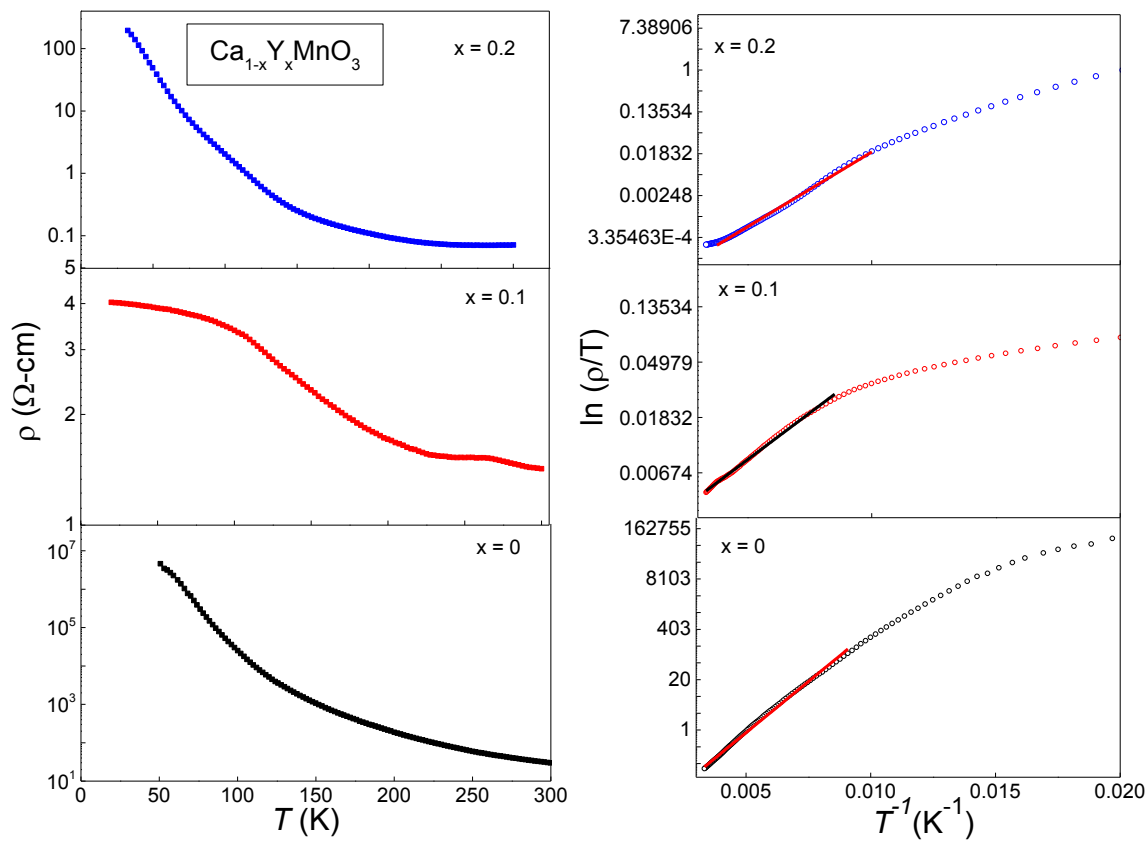

**Figure 3**

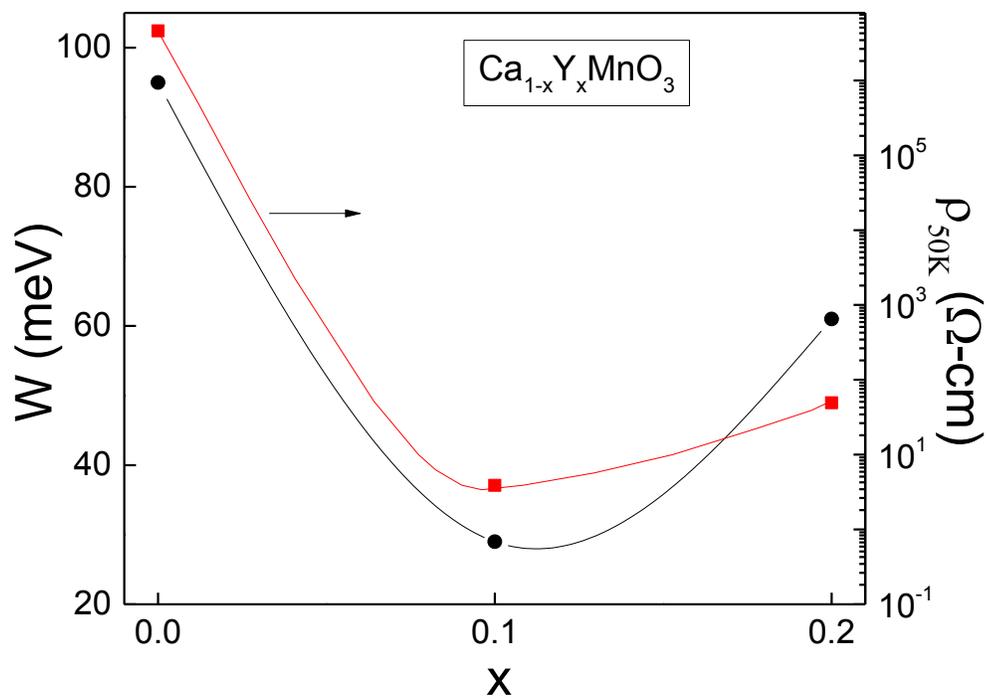

**Figure 4**

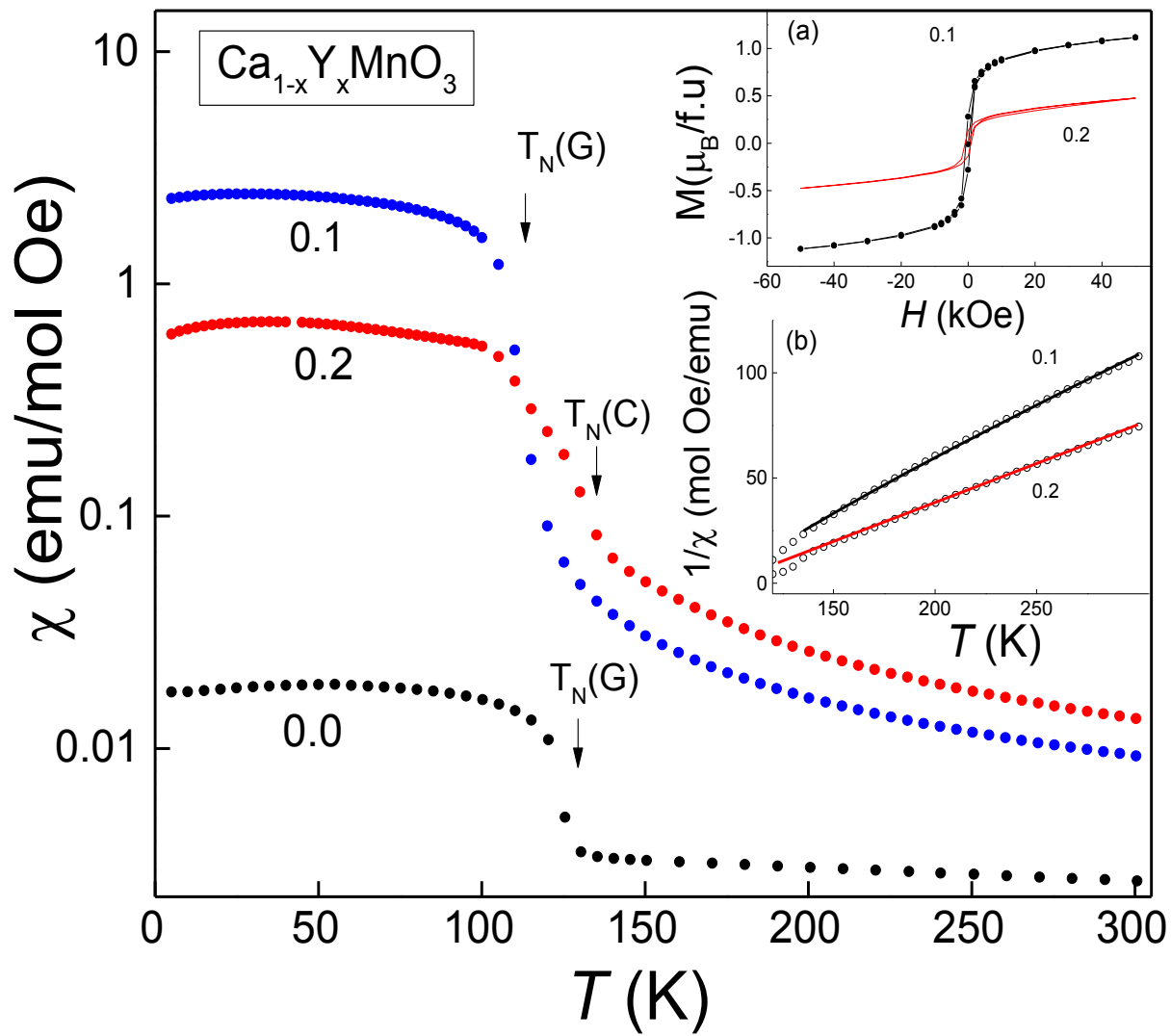

**Figure 5**

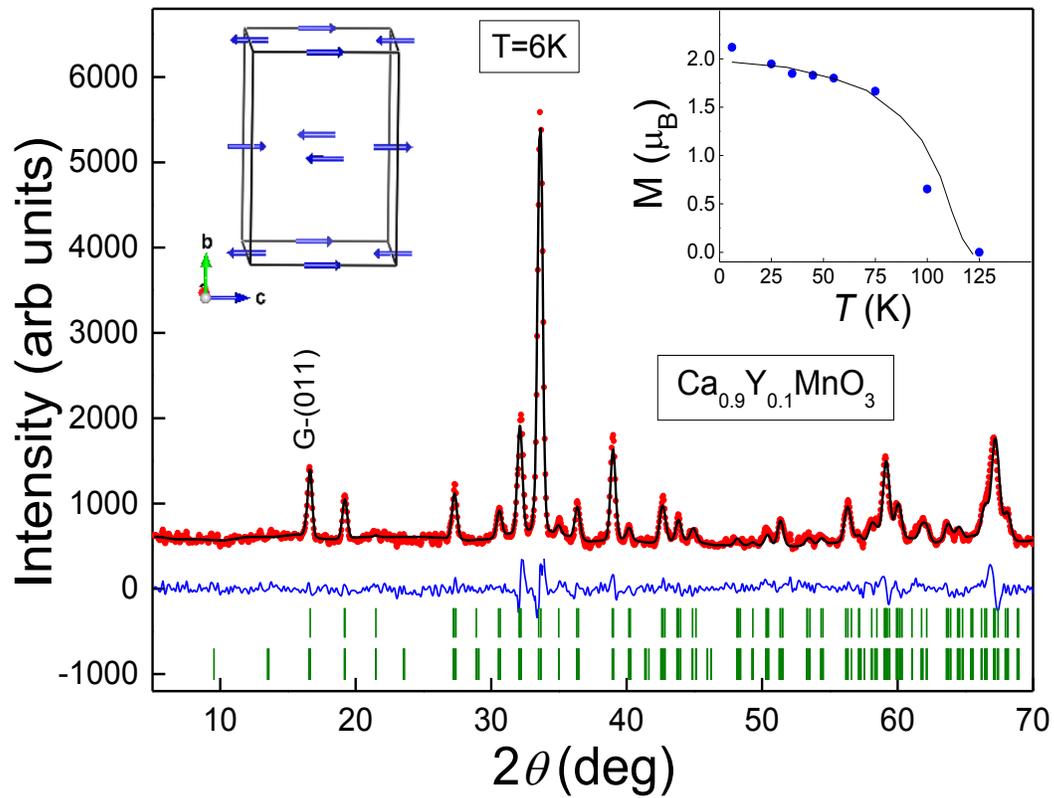

**Figure 6**

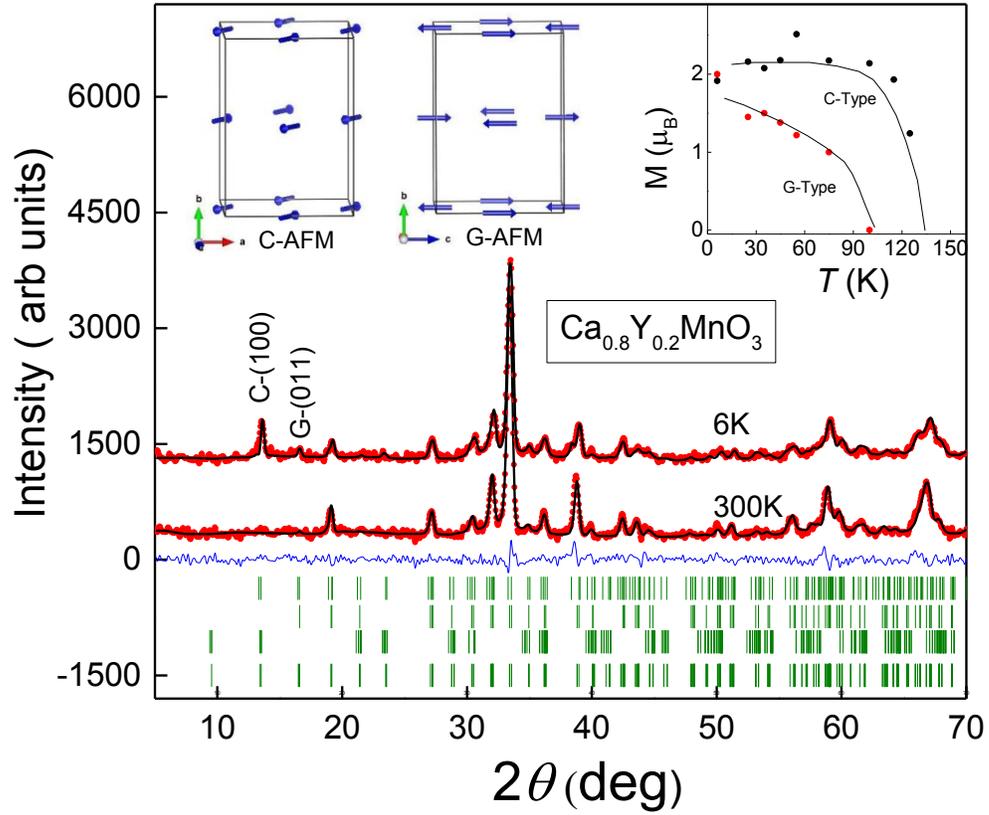

**Figure 7**

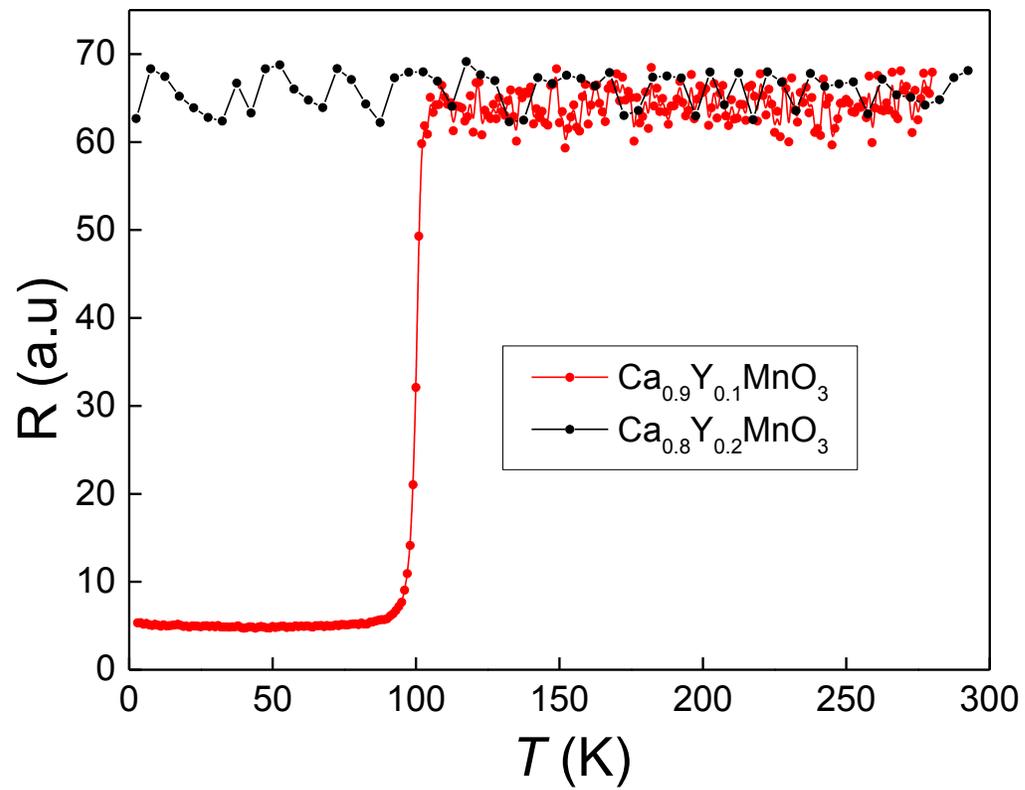

**Table 1** Results of Rietveld refinement of neutron diffraction pattern at 6 K, resistivity values, Curie – Weiss fit parameters, and variance ($\sigma^2$) for $Ca_{1-x}Y_xMnO_3$.

| | x = 0 [27] | x = 0.1 | x = 0.2 | |
|---|---|---|---|---|
| | | | Pnma (18%) | P2$_1$/m (82%) |
| a (Å) | 5.2771(10) | 5.2895 (8) | 5.3059 (3) | 5.3456 (2) |
| b (Å) | 7.4404(14) | 7.4572(14) | 7.4512 (14) | 7.4225 (20) |
| c (Å) | 5.2616(11) | 5.2508(8) | 5.2687 (9) | 5.2821(2) |
| Volume (Å$^3$) | 206.6 | 207.1 | 208.3 | 209.5 |
| β (°) | | | | 91.1(3) |
| $\rho_{300K}$ (Ω-cm) | 1.7 ± 0.4 | 1.4 ± 0.2 | 0.07 ± 0.01 | |
| θ (K) | -510 | -69 | -96 | |
| $\mu_{eff}$ ($\mu_B$) | 4.18 | 4.13 | 4.65 | |
| $\sigma^2 \times 10^{-3}$ | | 1 | 1.7 | |